\pdfoutput=1

\documentclass[%
 reprint,
 amsmath,amssymb,
 prl,superscriptaddress
]{revtex4-2}
\RequirePackage{xr-hyper}
\setcitestyle{super}
\usepackage{graphicx}
\usepackage{dcolumn}
\usepackage{bm,natbib}
\usepackage{xcolor}
\usepackage{comment}

\usepackage{hyperref}

\begin{document}

\preprint{APS/123-QED}


\title{On the interplay of electronic and lattice screening on exciton binding in\\ two-dimensional lead halide perovskites}

\author{Rohit Rana}
\affiliation{ 
Department of Chemistry, University of California, Berkeley, CA, USA, 94720
}%
\affiliation{Material Science Division Lawrence Berkeley National Laboratory, Berkeley, CA, USA, 94720
}
\author{David T. Limmer}%
 \email{dlimmer@berkeley.edu.}
\affiliation{ 
Department of Chemistry, University of California, Berkeley, CA, USA, 94720
}%
\affiliation{Material Science Division Lawrence Berkeley National Laboratory, Berkeley, CA, USA, 94720
}
\affiliation{Chemical Sciences Division, Lawrence Berkeley National Laboratory, Berkeley, CA, USA, 94720 \\
}
\affiliation{
Kavli Energy Nanoscience Institute, Berkeley, CA, USA, 94720
}

\date{\today}

\begin{abstract}
\noindent
\textbf{Abstract: }We use path integral Monte Carlo to study the energetics of excitons in layered, hybrid organic-inorganic perovskites in order to elucidate the relative contributions of dielectric confinement and electron-phonon coupling. While the dielectric mismatch between polar perovskite layers and non-polar ligand layers significantly increases the exciton binding energy relative to their three dimensional bulk crystal counterparts, formation of exciton polarons attenuates this effect.  The contribution from polaron formation is found to be a non-monotonic function of the lead halide layer thickness, which is clarified by a general variational theory. Accounting for both of these effects provides a description of exciton binding energies in good agreement with experimental measurements. By studying isolated layers and stacked layered crystals of various thicknesses, with ligands of varying polarity, we provide a systematic understanding of the excitonic behavior of this class of materials and how to engineer their photophysics.
\end{abstract}

\maketitle

A fundamental understanding of the photophysics of semiconducting materials is crucial for the engineering of efficient photovoltaic devices. One class of materials that has received significant recent attention are two-dimensional, layered lead halide perovskites due to their high operating efficiencies and robustness. \cite{li2022review,mao2018two,ghimire2021two}  
Two-dimensional materials generally exhibit large, tunable exciton binding energies due to their strong Coulomb interactions resulting from dielectric confinement.\cite{berkelbach2018optical,velicky2017two,chakraborty2022quantum} Work on bulk, three-dimensional perovskite lattices suggest that charge-phonon coupling and polaron formation underpin many of their observed electronic properties, including the low exciton binding energies.\cite{gan2021photophysics,ni2017real,straus2016direct,limmer2020photoinduced,park2022nonlocal}
\textcolor{black}{While the role of dielectric confinement has been studied for two-dimensional materials\cite{chakraborty2021dielectric,katan2019quantum,cho2021simulations,cho2019optical}, and the influence of polaronic physics has been probed extensively in three-dimensional perovskites\cite{park2022nonlocal,park2022renormalization,filip2021phonon}, it remains unclear how these two effects together influence the excited state properties of layered perovskites. To fill this knowledge gap, we study the interplay between dielectric confinement and charge-phonon interactions in layered perovskites using an imaginary time influence functional approach applied to a generalized Fröhlich model Hamiltonian.}
\cite{frohlich1954electrons,sio2022unified} We find that the incorporation of a dynamic lattice leads to a reduction in exciton binding energies, in opposition to the large increase due to dielectric confinement. Additionally, we find a non-monotonic dependence on the polaron binding energy as a function of lead halide layer thickness, and attribute it to an interplay between in-plane screening and polaronic localization.  

Three-dimensional lead halide perovskites have long charge carrier lifetimes and high power conversion efficiencies, but they suffer from stability problems that limit their use.\cite{shi2015low,lin2018perovskite,de2015impact,gangadharan2019searching} 
In contrast, two-dimensional layered lead halide perovskites, such as Ruddlesden-Popper phases, have improved shelf-life and tunability due to the addition of barrier ligand layers.\cite{liu2021research,leung2022stability,liu2019self} As a consequence they are actively investigated for optoelectronic applications. However, due to the heterogeneous composition of these layered materials,  their photophysics are different than their three-dimensional counterparts.\cite{liu2021research,blancon2018scaling,hansen2023mechanistic} Specifically, layered perovskites are quantum confined due to the small thickness of a typical inorganic layer relative to the excitonic radius.\cite{hanamura1988quantum} The dielectric contrast between the organic and inorganic layers leads to an enhancement of the electrostatic interaction between the electron and hole of an exciton.\cite{hanamura1988quantum,cho2019optical,cho2021simulations,blancon2018scaling} Moreover, excitonic properties are renormalized due to charge-phonon interactions arising from the soft inorganic lattice, and the interplay of these interactions together with quantum and dielectric confinement remain to be understood.\cite{hansen2023mechanistic,cho2019optical} 

\textcolor{black}{Computationally, the presence of exciton-phonon interactions leads to a variety of issues, such as the  breakdown of the Born-Oppenheimer approximation, that make it challenging to accurately solve for the electronic structure. One approach that has been taken to account for exciton-phonon interactions is the use of the semi-empirical Haken potential.\cite{haken1958theorie} However, this potential overestimates the exciton binding energy due to not properly capturing electron and hole polaron correlations.\cite{bajaj1974effect,movilla2023excitons} To remedy this, the Bajaj potential has introduced a phenomenological correction to the Haken potential\cite{bajaj1974effect}. Another approach is to introduce a phonon kernel within the GW/BSE framework.\cite{alvertis2024phonon,filip2021phonon,dai2024theory} While it is successful in predicting accurate exciton binding energies for various systems, GW/BSE is best suited for either handling the exciton-phonon interactions perturbatively or within the strong-coupling regime.} These considerations motivate the use of path integral methods\cite{feynman1962mobility,chandler1981exploiting,ceperley1995path,shumway2004quantum} 
which can incorporate electron-hole interactions and charge\textcolor{black}{-}lattice interactions at arbitrary strengths.\cite{park2022nonlocal,park2022renormalization,remsing2020effective} 

\begin{figure*}
\includegraphics[width=17.5cm]{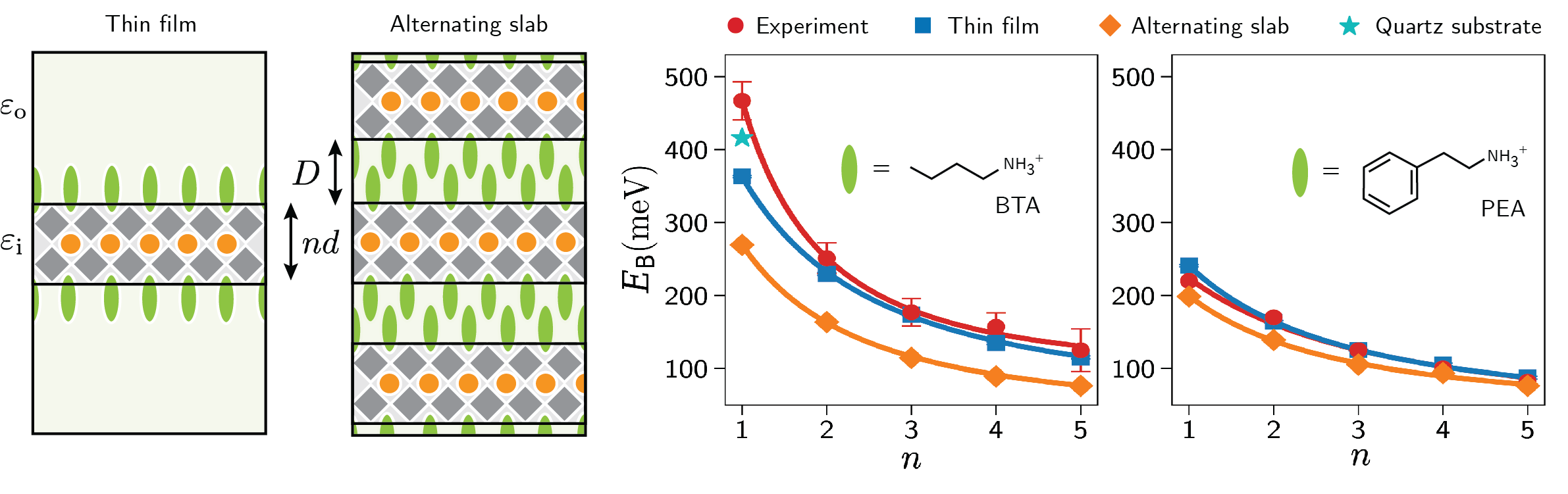}
\caption{\label{fig1}Thin film geometry and renormalized exciton binding energies at different quantum well thicknesses for (left) BTA and (right) PEA perovskites. The inorganic well is modeled with a film of thickness $nd$ and dielectric constant of $\varepsilon_\mathrm{i}$. The organic ligands are modeled with semi-infinite thickness and a dielectric constant of $\varepsilon_\mathrm{o}$. Circle markers represent experimental values obtained from literature. Square and diamond markers represent calculations done in this work for thin film and crystal geometries, respectively. Solid lines are fits to Eq. \ref{scaling}.}
\end{figure*}

In this work, we employ an imaginary time influence functional approach to the thermal path integral to study a continuum model of electron and hole quasiparticles interacting with the layered perovskite lattice.\cite{park2022nonlocal,park2022renormalization,park2023biexcitons} 
As depicted in Fig. \ref{fig1}, we apply this approach to two model systems with organic ligands,  n-butylammonium (BTA) and phenylethylammonium (PEA), of varying polarity and having stoichiometry (BTA)$_2$(MA)$_{n-1}$Pb$_n$I$_{3n+1}$ and (PEA)$_2$(MA)$_{n-1}$Pb$_n$I$_{3n+1}$, where $n$ is the number of inorganic sublayers. Modifying $n$ changes the quantum well thickness and consequently, the degree of confinement.\cite{blancon2018scaling,hansen2023mechanistic,cho2019optical,cho2021simulations}

The model Hamiltonian we consider is defined as
\begin{equation} \label{totalH}
\hat{\mathcal{H}} = \hat{\mathcal{H}}_{\text{eh}} + \hat{\mathcal{H}}_{\text{ph}} + \hat{\mathcal{H}}_{\text{int}}
\end{equation}
where $\hat{\mathcal{H}}_{\text{eh}}$ corresponds to the electronic degrees of freedom, $\hat{\mathcal{H}}_{\text{ph}}$ corresponds to the phonon degrees of freedom, and $\hat{\mathcal{H}}_{\text{int}}$ corresponds to the interaction between the phonon and electronic degrees of freedom. We employ an effective mass approximation and define the electronic Hamiltonian as
\begin{equation} \label{elecH}
\hat{\mathcal{H}}_{\text{eh}} = \frac{\hat{\boldsymbol{p}}_{\text{e}}^2}{2m_e} + \frac{\hat{\boldsymbol{p}}_{\text{h}}^2}{2m_h} + \hat{V}(|\hat{\boldsymbol{r}}_{\text{e}} - \hat{\boldsymbol{r}}_{\text{h}}|) 
\end{equation}
where the subscript $e$ and $h$ indicates electron and hole, $\hat{\boldsymbol{p}}$ and $\hat{\boldsymbol{r}}$ are the in-plane momentum and position, and  $m_e = m_h = 0.20$\cite{cho2019optical} are the effective quasiparticle masses in units of the bare electron mass. We use continuum electrostatics and define $\hat{V}(|\hat{\boldsymbol{r}}_{\text{e}} - \hat{\boldsymbol{r}}_{\text{h}}|)$ to be the electron-hole interaction at the equilibrium lattice geometry. 

Layered perovskites are dielectrically inhomogeneous, with planar interfaces in the stacking direction separating polarizable inorganic layers which host optical excitations, and less polar ligand layers which act as barriers. As a consequence, the effective interaction between electrons and holes is more complicated than its three-dimensional, homogeneous dielectric equivalent. In a cylindrical coordinate system, the electrostatic potential at position $\mathbf{s} = (\mathbf{r},z)$ created by an electron in the inorganic well layer at position $\mathbf{s}' = (\mathbf{r}',z')$ satisfies Poisson's equation. Within the inorganic region hosting the point charge at $(\mathbf{r}=0, z)$, 
\begin{equation} \label{poiss}
\nabla^2 V(\mathbf{r'},z',z)= -\frac{4 \pi e}{\varepsilon_{i}}\delta(\mathbf{r})\delta(z'-z) \quad z,z'\leq nd
\end{equation}
where $e$ is the fundamental charge, $\epsilon_\mathrm{i}$ is the inorganic well layer dielectric constant, and $d$ the thickness of a single inorganic layer. Outside of the first inorganic layer, the electrostatic potential satisfies $\nabla^2 V(\mathbf{r}', z',z)= 0$, and the full solution must satisfy the dielectric discontinuity boundary conditions 
\begin{equation}
\varepsilon_\mathrm{i}\nabla V(\mathbf{r}', k nd+l D,z)= \varepsilon_\mathrm{o}\nabla V(\mathbf{r}', k nd+l D,z) 
\end{equation}
where $\varepsilon_\mathrm{o}$ in an organic layer dielectric constant, $D$ is the width of the organic layer, and $k,l$ index the locations of the dielectric boundaries. Throughout this work, the length of one inorganic sublayer $d$ is $6.39 \text{\AA}$.\cite{oku2015crystal,cho2019optical,cho2021simulations}

We consider two distinct experimentally relevant dielectric geometries, as depicted in Fig. \ref{fig1}. The first is a thin film of inorganic material surrounded by two semi-infinite surrounding organic layers, in which case $k=\{0,1\}$ and the $lD$ term is ignored.\cite{barrera101978point} In this case, the semi-infinite organic regions model exfoliated perovskites dispersed in organic solvents.\cite{yaffe2015excitons} This is equivalent to the Rytova-Keldysh theory.\cite{rytova2020screenedpotentialpointcharge,keldysh2024coulomb} 
The second geometry is a crystal or infinitely-alternating slab geometry\cite{muljarov1995excitons,guseinov1984coulomb} where a unit cell consists of an inorganic layer of length $nd$, and organic layer of length $D = 8.81  \text{\AA, }9.92 \text{\AA}$ for BTA and PEA, respectively.\cite{muljarov1995excitons} In this case, $k,l=\{-\infty,\infty\}$. For each geometry, we use analytic representations of the solutions of Poisson's equations in a Fourier-Bessel basis and then numerically inverse Fourier transform them to evaluate the electrostatic energy in our Monte Carlo simulations (SI). Additionally, since the exciton binding energies of the model systems have been measured to be much bigger than the dominant longitudinal optical mode phonon energies,\cite{blancon2018scaling} we use high-frequency dielectric constants to parameterize the potential energy functions for both geometries. Specifically, the inorganic layers have a dielectric constant of $\varepsilon_\text{i} = 6.1$.\cite{ishihara1989exciton} For the organic layers, we use $\varepsilon_\text{o} = 2.1$\cite{ishihara1990optical} and $\varepsilon_\text{o} = 3.3$\cite{hong1992dielectric} for the BTA and PEA ligands, respectively.

Using the solution to Poisson's equation for either geometry, the electron-hole interaction in Eq. \ref{elecH} is
\begin{flalign} \label{pot}
V(r) = -e\int dz_h\int  dz_e \, P(z_e) P(z_h) V(r,z_e,z_h)  
\end{flalign}
where $r = |\boldsymbol{r}_{\text{e}} - \boldsymbol{r}_{\text{h}}|$ is the in-plane distance between the electron and hole, $z_e$ is the $z$ coordinate of the electron, $z_h$ is the $z$ coordinate of the hole, and $P(z_e)$ and $P(z_h)$ are the marginalized probability densities for the electron and hole in $z$. 
Without loss of generality, we average out the $z$-dependencies in Eq. \ref{pot} using particle-in-a-box 
ground state probability densities for $P(z_e)$ and $P(z_h)$ for a box size of length $nd$. We find that the exciton binding energy does not change by more than a few meV if uniform probability densities are used instead. This observation is consistent with previous work.\cite{hansen2023mechanistic} 

We are interested in low energy excitations and low temperatures, so to incorporate the lattice in our model Hamiltonian, we assume that it is described by harmonic modes
\begin{equation} \label{phH}
\hat{\mathcal{H}}_{\text{ph}} = \frac{1}{2}\sum_{\boldsymbol{k}}(\hat{\boldsymbol{p}}_{\boldsymbol{k}}^2 + \hat{\omega}_{\boldsymbol{k}}^2 \hat{\boldsymbol{q}}_{\boldsymbol{k}}^2)
\end{equation}
where $\hat{\boldsymbol{p}}_{\boldsymbol{k}}$ and $\hat{\boldsymbol{q}}_{\boldsymbol{k}}$ are mass-weighted, in-plane momentum and position coordinates of a phonon described by a wavevector of  $\boldsymbol{k}$. We consider only long-wavelength longitudinal optical modes, treated as dispersionless $\omega_{\boldsymbol{k}} \approx \omega_{\text{LO}}$, where $\omega_{\text{LO}}$ corresponds to the longitudinal optical mode frequency,  with \textcolor{black}{$\hbar\omega_{\text{LO}} = 12.4 \text{ meV}$ for the BTA perovskites\cite{ni2017real} and $\hbar\omega_{\text{LO}} = 14.0 \text{ meV}$ for the PEA perovskites.}\cite{straus2016direct} We assume the charge-phonon interaction is bilinear, which allows us to write\cite{feynman2018statistical,park2022nonlocal,park2022renormalization}
\begin{equation} \label{frohH}
\hat{\mathcal{H}}_{\text{int}} =  \sqrt{\frac{2\omega_{\text{LO}}}{\hbar}} \sum_{\mathbf{k}} \hat{q}_{\mathbf{k}} \bigg[ |g(\mathbf{k})|e^{i\mathbf{k} \cdot \hat{\mathbf{r}}_{\text{e}} } - |g(\mathbf{k})|e^{i\mathbf{k} \cdot \hat{\mathbf{r}}_{\text{h}} }  \bigg]  
\end{equation}
where $|g(\mathbf{k})|$ represents the magnitude of the  interaction strength and $\hbar$ is the reduced Planck's constant. This form is sufficient to capture the interaction between each charge and the induced dipole of the lattice. The Hamiltonian in Eq. \ref{frohH} could also be extended to higher temperatures using effective phonon theories.\cite{park2022nonlocal}

The specific model for the charge-lattice coupling we consider is a generalized Fröhlich model derived for a charge screened by a longitudinal phonon in a thin film surrounded by two semi-infinite non-polarizable layers.\cite{sohier2016two,sio2022unified} This yields a charge-phonon interaction strength of 
\begin{subequations} \label{ephMatrix}
\begin{flalign}
|g(\textbf{k})| = \frac{ \bigg[\pi e^2 nd \hbar \omega_{\text{LO}} (\varepsilon_\mathrm{i}^0 - \varepsilon_\mathrm{i}) /2A \bigg]^{1/2}} {\varepsilon_\text{o} +n  r_{\text{eff}}k}  
\end{flalign}
where
\begin{flalign}
r_{\text{eff}} = \varepsilon_\text{o} d \bigg[ \frac{\varepsilon_\text{o}}{3 \varepsilon_\text{i}} + \frac{\varepsilon_\text{i}}{2 \varepsilon_\text{o}} \bigg( 1 - \bigg[\frac{\varepsilon_\text{o}}{\varepsilon_\text{i}}\bigg]^2 \bigg) \bigg]
\end{flalign}
\end{subequations}
is a dielectric screening length, $A$ is the area of the unit cell, and $k = |\mathbf{k}|$.
Throughout this work, we use a static dielectric constant of $\varepsilon_\mathrm{i}^0 = 13.0$.\cite{ishihara1989exciton,ishihara1990optical} We note that when compared to three dimensional materials, the value of $(\varepsilon_\mathrm{i}^0 - \varepsilon_\mathrm{i})$ is smaller for two dimensional materials due to the presence of fewer polarizable inorganic layers.\cite{sio2023polarons} \textcolor{black}{Nonetheless, we find that the model Hamiltonian is sensitive to the choice of $\varepsilon_\mathrm{i}^0$ (SI).}\textcolor{black}{While we use fixed LO phonon frequencies and dielectric constants for this work, these assumptions can be relaxed \textcolor{black}{for different values of $n$}, and the corresponding changes to the charge-phonon interaction can be understood from Eq. \ref{ephMatrix}.}

\textcolor{black}{For the path integral Monte Carlo method, we can write the action corresponding to the model Hamiltonian in Eq. \ref{totalH} within an imaginary time interval and 
uniformly discretize it into $N$ slices (SI).\cite{park2022renormalization}} Then, we utilize the Metropolis-Hastings algorithm\cite{metropolis1953equation,hastings1970monte} with proposed uniform displacements both of individual slices and the center of mass within a periodic two-dimensional box of 500 $\times$ 500 a.u.  
With path integral Monte Carlo, we can compute the exciton binding energy from
\begin{equation} \label{excBinding}
 E_\text{B} =  \lim_{T\to 0}  \langle E_{\text{eh}}\rangle -  \langle E_{\text{e}}\rangle  -  \langle E_{\text{h}} \rangle
\end{equation}
where $E_{i}$ is a simulation with ($i=\mathrm{eh}$) an electron and hole, or ($i=\mathrm{e}$) just an electron or ($i=\mathrm{h}$) hole, and brackets denote thermal average.
We use the virial estimator\cite{herman1982path} to obtain values of the average energy at finite temperatures ranging from 50 K to 400 K. To extrapolate the ground-state exciton binding energy to 0 K, we use a two-level system fit 
in accordance with Boltzmann statistics that is parameterized with the ground-state exciton binding energy and the difference between the two energy levels. The large binding energies make this especially convenient. We used between 2 and 5 million Monte Carlo sweeps for each temperature to estimate averages.

We compute the exciton binding energy with the Hamiltonian defined in Eq. \ref{totalH} as a function of increasing inorganic layer thickness, $n$. This yields insight into how quantum confinement, dielectric confinement, and electrostatic effects arising from the choice of geometry affect exciton binding energies. We find for the thinnest materials, the exciton binding energies are nearly an order of magnitude larger than their bulk three dimensional counterparts.\cite{park2022nonlocal} However, as shown in Fig. \ref{fig1}, we see that for larger values of $n$, there is a dramatic reduction in the exciton binding energy. \textcolor{black}{Consistent with previous studies \cite{chakraborty2021dielectric,katan2019quantum,cho2021simulations,cho2019optical}}, these trends are a consequence of two factors. The first is that quantum confinement decreases for larger values of $n$, generally allowing charges to become more diffuse, weakening their interactions. The second is that dielectric confinement also decreases for larger values of $n$, so that the electron-hole interaction is more strongly screened by the more polarizable inorganic layer. 

\begin{figure}[t]
\includegraphics[width=8.5cm]{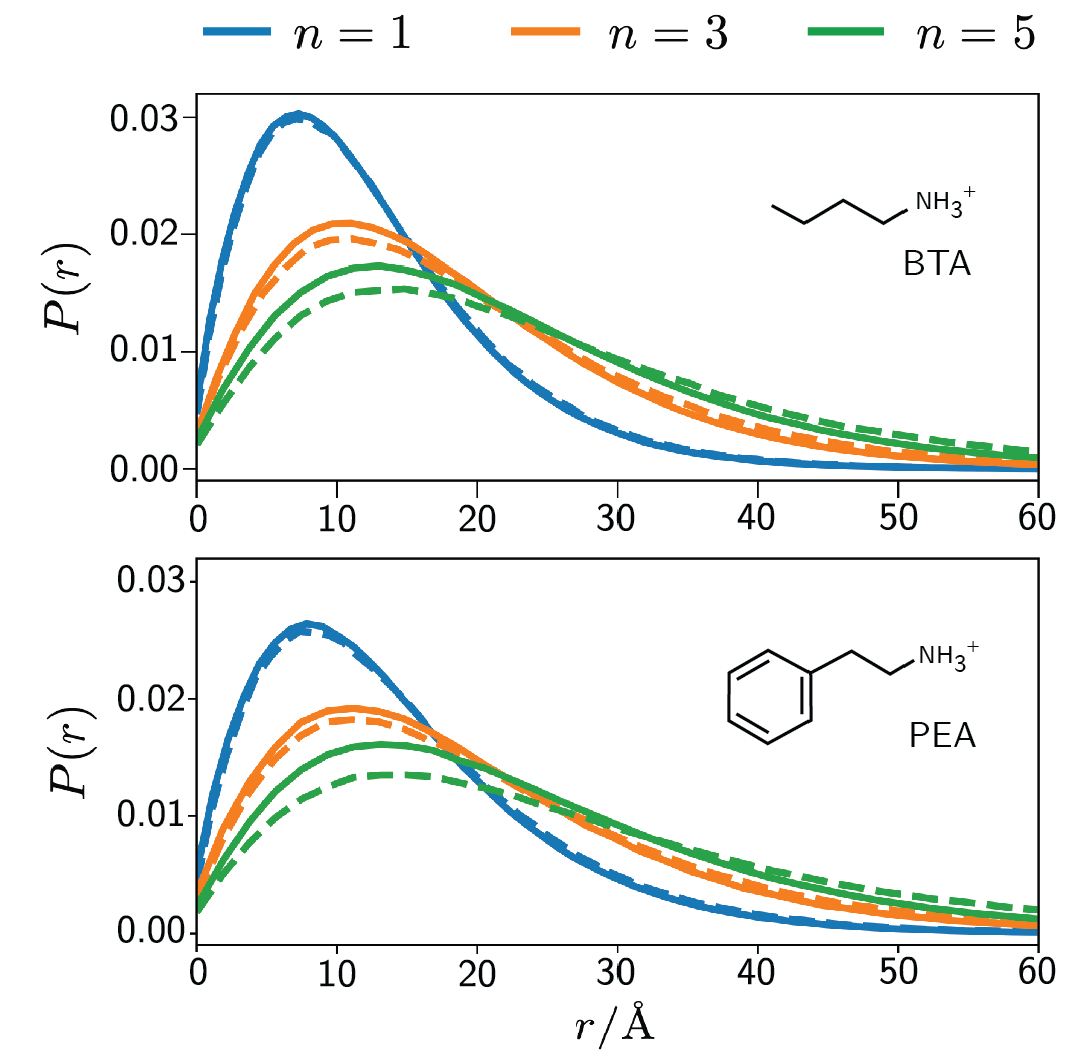}
\caption{\label{fig3} Exciton probability densities of (top) BTA perovskites and (bottom) PEA perovskites as a function of quantum well thickness. Solid lines represent a bare exciton, and dashed lines account for exciton polaron formation.}
\end{figure}

The role of quantum confinement can be assessed by considering the size of the exciton relative to the layer thickness. The size of the exciton is determined from its pair distribution function,
\begin{equation}
P(r) = \langle \delta (r-|\mathbf{r}_e-\mathbf{r}_h|)\rangle 
\end{equation}
in the low temperature limit. 
Figure \ref{fig3} shows $P(r)$ computed for the exfoliated slab geometry. We find the exciton becomes more delocalized as $n$ increases for both ligand systems. From $n=1$ to $n=5$, the exciton size, defined as the peak in $P(r)$, increases by nearly a factor of 2. This is a consequence of the reduced dielectric confinement for larger values of $n$. The size dependence is consistent with experimental observations that found a reduction of oscillator strength with increasing $n$.
\cite{yaffe2015excitons}  Although not shown, the same trends are obtained using the alternating slab geometry.

To understand the role of dielectric confinement, we compare the exciton binding energies for the two different ligand systems in Fig. \ref{fig1} for a given geometry at a given value of $n$. We find the exciton binding energies of the PEA perovskites are smaller than the BTA perovskites. This is a result of the ligand dielectric environment becoming more similar to the inorganic perovskite layers, since the aromatic group on PEA is more polarizable. The reduced dielectric confinement in the PEA perovskites and their concomitant reduced binding energies also explains a previous observation showing that these materials have a higher photoluminscence quantum yield in comparison to the BTA perovskites.\cite{quan2021vibrational} For a given layered perovskite at a given value of $n$, Fig. \ref{fig1} shows that the exciton binding energy is smaller for the alternating slab or crystal geometry than the film geometry. This is a result of a reduction in dielectric confinement due to an exciton in one inorganic layer being screened by other surrounding inorganic layers in the crystal geometry. This interaction is not present for the film geometry since it is assumed to be surrounded by organic solvent. Similar to the comparison of PEA perovskites to BTA perovskites, we expect that crystalline layered perovskites have longer radiative lifetimes and higher photoluminscence quantum yields than their exfoliated counterparts.

Dielectric confinement can be modeled by an effective fractional-dimension scaling law\cite{he1991excitons,mathieu1992simple,cho2021simulations,blancon2018scaling}
\begin{equation} \label{scaling}
E_{B} =  \frac{E_{\text{3D}}}{\big[ 1 - \frac{\Omega}{2} \big]^2}
\end{equation}
where \textcolor{black}{$\Omega = \kappa \exp[-nd/(2r_{\text{3D}})]$}  is the difference in dimensionality from three, $E_\text{3D}$ is the Rydberg energy, $nd$ is the quantum well thickness, $r_{\text{3D}}$ is the three-dimensional Bohr radius, and $\kappa$ is a fitting parameter.
For simplicity, we first use exciton binding energies obtained with simulations in the absence of phonons (Eq. \ref{elecH}) along with $E_\text{3D} = 37 \text{ meV}$ and $r_{3D} = 32 \text{ \AA}$.\cite{cho2021simulations} For all of the layered perovskites studied in this work, we find $\Omega$ to be smaller than $2$ for the $n = 1$ case. This exemplifies that quantum confinement in the $z$-direction is complemented with dielectric effects from the heterogeneous environment. Additionally, we find $\kappa$ to take a range of $1.34$ to $1.55$. The perovskites with a larger value of $\kappa$ will exhibit stronger dielectric confinement than those with smaller values of $\kappa$. This implies that two-dimensional systems with greater dielectric mismatch will require thicker layers to approach the bulk crystal regime than those with smaller mismatches. We further assess the utility of the scaling relation by parameterizing Eq. \ref{scaling} with exciton binding energies obtained in the presence of a dynamic lattice (Eq. \ref{totalH}) and allowing $E_{\text{3D}}$, $r_{\text{3D}}$, and $\kappa$ to all be fitting parameters. The resulting fits are shown as solid lines in Fig.~\ref{fig1} and model the data quite well. We also find the fitted bulk exciton binding energies and bulk exciton radii to be in good agreement with experiment.\cite{baranowski2020excitons}

The simulations shown in Fig. \ref{fig1} also highlight that the minimal ingredients in our model are sufficient to predict theoretical exciton binding energies in good agreement with experiment. We find that the exciton binding energies obtained with the thin film geometry are in good agreement with experimental measurements of BTA exfoliated perovskites.\cite{blancon2018scaling} Figure \ref{fig1} shows that the renormalized exciton binding energies obtained with the crystal geometry also shows good agreement with experimental values of PEA perovskite crystals. \cite{delport2019exciton,urban2020revealing} 
The only significant quantitative disagreement is the case of the $n = 1$ film geometry calculation in the BTA perovskite. However, we note that in the experimental setup, the exfoliated perovskites are placed on top of quartz substrates and are surrounded by vacuum for measurement of the binding energies. Therefore the simplifying assumption of being dispersed in a low dielectric solvent is not valid. A better description for this case is to employ a generalized film model where the layers surrounding the film on the top and bottom are different and the thin film is a composite dielectric.\cite{barrera101978point} To set up such a model we replace the inorganic layer with a thickness weighted dielectric constant \cite{yaffe2015excitons,muljarov1995excitons} 
\begin{equation} \label{averagedDielec}
(\varepsilon_\text{i})_{\text{avg}} = \frac{\varepsilon_\mathrm{i}' nd + \varepsilon_\mathrm{o}' D}{nd + D}
\end{equation}
for a film of length $nd$, where $\varepsilon_\mathrm{i}'$ and $\varepsilon_\mathrm{o}'$ are chosen to be 6.1 and 2.1, respectively. This thin film is placed in contact with a semi-infinite top surrounding layer of air with a dielectric constant of $1.0$, and a semi-infinite bottom layer is chosen to be quartz with a dielectric constant of $3.9$.\cite{zahra2017comparative}  With this model, we find the $n = 1$ renormalized exciton binding energy to be 416 meV, which is closer to the experimental value. For larger values of $n$, we find no significant differences in comparison to the exciton binding energies plotted in Fig. \ref{fig1}. This indicates that the role of the dielectric environment surrounding the inorganic perovskite layer becomes negligible for larger values of $n$.

\begin{figure}[t]
\includegraphics[width=8.5cm]{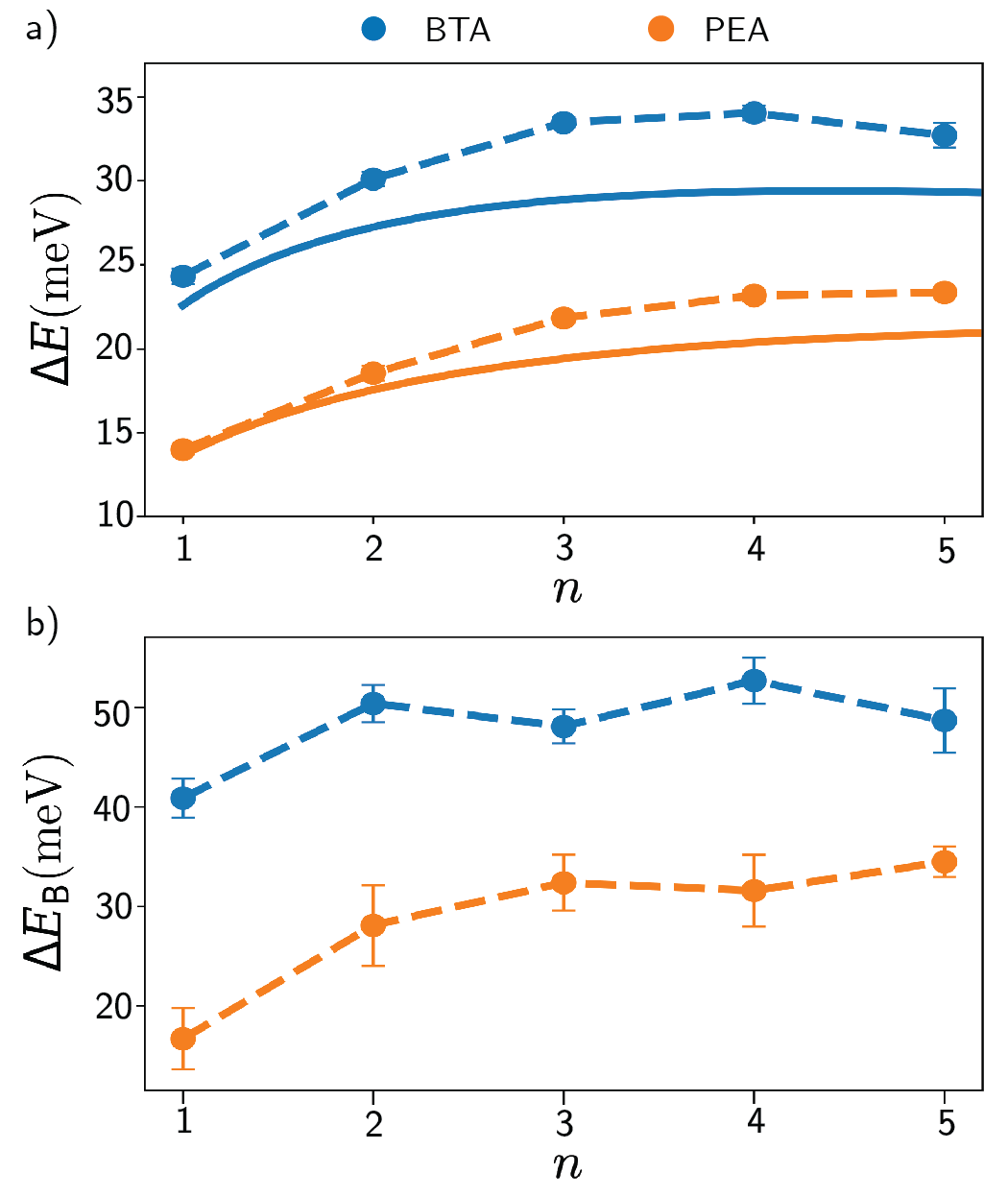}
\caption{\label{fig2} a) Polaron binding energies  and b) renormalization to exciton binding energies as a function of quantum well thickness for BTA and PEA perovskites. \textcolor{black}{Solid lines represent variational calculations.} }
\end{figure}

The good quantitative agreement between theoretical exciton binding energies and experimental values is due in large part to the renormalization of the binding energy from polaron formation. This is demonstrated by computing the change in energy for a single electron (or hole) due to interactions with phonons. This quantity, also known as the polaron binding energy, is defined as 
\begin{equation} \label{polaronBinding}
\Delta E = | \langle E_{\text{e}} \rangle_{\alpha \neq 0 } - \langle E_{\text{e}} \rangle_{\alpha = 0}|
\end{equation}
for a dimensionless electron-phonon coupling parameter $\alpha$ and extracted in the low temperature limit. From the data in Fig. ~\ref{fig2} a), \textcolor{black}{we find that the polaron binding energies are larger than twice the three-dimensional values in magnitude \cite{park2022renormalization}, thus highlighting that the heterogeneous dielectric environment in two-dimensional perovskites stabilizes the free charges more than expected from a uniform dielectric system.} Additionally,  the polaron binding energy is larger in magnitude for the BTA perovskites in comparison to the PEA perovskites. Since the longitudinal optical mode frequencies and effective masses are similar in value for both perovskites, this larger polaron binding is a result of stronger dielectric confinement in the BTA system versus the PEA system. From Eq.~\ref{excBinding} the exciton binding energy is the difference of the exciton energy and the free charges, so polaron formation reduces the exciton binding energy. 
 
 An offsetting factor to polaron formation is the stabilization of the exciton energy through exciton polaron formation. The renormalization to the exciton binding energy due to charge-phonon interactions is computed from
 \begin{equation} \label{excitonRenorm}
 \Delta E_{\text{B}} = |( E_{\text{B}})_{\alpha \neq 0} - ( E_{\text{B}})_{\alpha = 0}|
 \end{equation}
 in the low temperature limit and is shown in Fig.~\ref{fig2}b). Our exciton polaron binding energies are also significant \textcolor{black}{when compared to the bulk value\cite{park2022renormalization}. However, these values are smaller than twice the respective single charge polaron binding energies, resulting in a net destabilization of the exciton.} \textcolor{black}{Furthermore, there is a weak dependence of the exciton polaron binding energy with respect to $n$, a trend also observed in a Haken potential\cite{movilla2023excitons}.} We notice from Fig. \ref{fig3} that for larger values of $n$ where the renormalization to exciton binding is larger, the exciton becomes slightly more delocalized in comparison to the bare exciton. 
 The $n$ dependence can be understood by considering an exciton as a dipole. From Fig.~\ref{fig3}, we observe that the increasing size of the exciton implies that the excitonic dipole moment increases as a function of $n$. For small values of $n$ where the exciton has a smaller dipole moment, the exciton interacts more weakly with the polarization generated by the inorganic lattice.
 
 %
%




Our calculations in Fig.~\ref{fig2}a) show a non-monotonic change in polaron binding energy as function of $n$ for both perovskites. In order to interpret this unanticipated dependence on layer thickness, we consider a weak coupling regime \textcolor{black}{and derive a variational bound (SI)}
\begin{equation} \label{bound}
\Delta E \geq  \alpha  \hbar \omega_{\text{LO}} \, \xi \left ( \frac{n r_{\text{eff}}}{\varepsilon_\text{o}} \Bigg/\sqrt{\frac{\hbar}{2 m_e \omega_{\text{LO} }}} \right )
\end{equation}
where \textcolor{black}{$\alpha$ is a dimensionless coupling constant that is inversely proportional to the layer thickness and}
the dimensionless scaling function, $\xi$, 
\begin{equation} \label{s}
\xi(a) = 4 \frac{ (a^4-a^2)       \ln \big(a\big) + \pi a^3  - a^2 - a^4          }{ (1 + a^2)^2}
\end{equation}
depends on the ratio of the dielectric confinement length $n r_{\text{eff}}/\varepsilon_\text{o}$ to the polaron radius $\sqrt{\hbar/2 m_e \omega_{\text{LO} }}$.  The non-monotonic behavior is captured by the bound, as $\alpha$ decreases with layer thickness, while $\xi$ is an increasing function of layer thickness. The bound \textcolor{black}{ underestimates} the binding energy slightly, and it is not easily extendable to an equivalent bound for the exciton polaron energy.

A maximum is predicted where the two lengths are equal, also consistent with our calculations.  Physically, the maximum in polaron binding results from two competing mechanisms. First, as $n$ increases, there is more polarizable material present in the film layer. From this, we expect the electron-phonon interaction to increase\cite{sio2023polarons} as a function of $n$. However, as $n$ increases, it is easier to form a dipole in the lattice because it is better screened by the surrounding environment. As a consequence the effective dipole strength decreases, in turn also decreasing the reaction field from the charge and causing the stabilization of the polaron to lessen. 

\textcolor{black}{We see that from adding the values of Fig.~\ref{fig1} and Fig.~\ref{fig2}b), we would consistently overestimate the experimental exciton binding energies if we exclude polar lattice vibrations.} By accounting for dynamical screening arising from charge-phonon interactions, we find exciton binding energies of two model layered perovskites to be in good agreement with experiment. Using a Fröhlich-like model Hamiltonian with a path integral Monte Carlo framework  allows us to capture higher levels of \textcolor{black}{exciton-phonon} correlations not easily obtained in electronic structure methods like GW/BSE.\cite{rohlfing2000electron} 
The complex interplay between quantum confinement, dielectric confinement, and charge-phonon interactions can be analyzed through the framework established in this work. Our approach suggests a promising avenue towards the design and understanding of photovoltaic devices. \textcolor{black}{This includes further understanding the influence of interlayer phonons in 2D heterostructures\cite{lee2024phonon} or how exciton-phonon coupling behaves in Moiré superlattices.}
\newline
\indent
\textbf{Supporting Information.} Includes the derivation of the effective electron-hole interaction, additional simulation details, the derivation of the variational bound within a path integral framework, a table of electrostatic potential parameters, and a figure of the variational bound as a function of quantum well thickness

\begin{acknowledgments}
\textbf{Acknowledgments.} This work was supported by the US Department of Energy,
Office of Science, Office of Basic Energy Sciences, Materials Sciences and Engineering Division, under Contract No. DE-AC02-05CH11231 within the Fundamentals of Semiconductor Nanowire Program (KCPY23).  D.T.L. acknowledges the Alfred P. Sloan Foundation. The authors thank Dr. Eric R. Heller and Dr. Jorge L. Rosa-Raíces for useful discussions about the path-integral framework.\end{acknowledgments}

\textbf{References}
\bibliography{CITE_main}
\end{document}


\def\thesubsubsection{\arabic{subsubsection}}

\title{Supplementary Material for ``On the interplay of electronic and lattice screening on exciton binding in
two-dimensional lead halide perovskites"}

\author{Rohit Rana}
\affiliation{ 
Department of Chemistry, University of California, Berkeley, CA, USA, 94720
}%
\affiliation{Material Science Division Lawrence Berkeley National Laboratory, Berkeley, CA, USA, 94720
}
\author{David T. Limmer}%
 \email{dlimmer@berkeley.edu.}
\affiliation{ 
Department of Chemistry, University of California, Berkeley, CA, USA, 94720
}%
\affiliation{Material Science Division Lawrence Berkeley National Laboratory, Berkeley, CA, USA, 94720
}
\affiliation{Chemical Sciences Division, Lawrence Berkeley National Laboratory, Berkeley, CA, USA, 94720 \\
}
\affiliation{
Kavli Energy Nanoscience Institute, Berkeley, CA, USA, 94720
}

\date{\today}

\maketitle

\renewcommand{\theequation}{S.\arabic{equation}}
\renewcommand{\thetable}{S\arabic{table}.}
\renewcommand{\thefigure}{S\arabic{figure}}

\section{Derivation of Effective Electron-Hole Interaction}
\subsection{Introduction}
\noindent
Our goal is to find the electron-phonon matrix element $|g(\mathbf{k})|$ for a Rytova-Keldysh film geometry. We will focus on longitudinal optical (LO) phonons. The polarization generated by an atomic displacement corresponding to a LO phonon with in-plane momentum $\mathbf{k}$ is given  as \cite{sohier2016two}
\begin{equation} \label{s1}
\mathcal{P}(r,z) = \frac{e^2}{A} \sum_{a} \frac{Z_a \cdot (\mathbf{k}^a_{\text{LO}}/|\mathbf{k}^a_{\text{LO}}|) }{\sqrt{2M_a \omega_{\text{LO}}/\hbar}}{p}(z)e^{i \mathbf{k} \cdot \mathbf{r}}  
\end{equation}
where the summation runs over the atoms of the unit cell. Here, $Z_a$ is the Born effective charge tensor, $A$ is the area of the unit cell, $p(z)$ is the z-direction polarization profile, $e$ is the fundamental charge, $\hbar$ is the reduced Planck's constant, $\omega_{\text{LO}}$ is the LO phonon frequency, and $M_a$ is the mass of atom $a$. 
Without loss of generality, assume that the dielectric tensor is isotropic, and assume that the composition of the two layers surrounding the film layer are the same. Then, letting $k = |\mathbf{k}|$ we can write (see Appendix A of \cite{sohier2016two})
\begin{subequations} \label{s2}
\begin{flalign}
|g(\mathbf{k})| = \frac{C}{\varepsilon_{\text{i}}} \bigg[ \frac{2}{knd} \bigg(1 + \frac{e^{-knd}-1}{knd} \bigg) + \frac{2(1-e^{-knd})^2 (\delta + \delta^2 e^{-knd})}{(knd)^2 (1-\delta^2 e^{-2knd})}  \bigg]
\end{flalign}
where 
\begin{flalign}
\delta = \frac{\varepsilon_{\text{i}} - \varepsilon_{\text{o}}}{\varepsilon_{\text{i}} + \varepsilon_{\text{o}}} 
\end{flalign}
and
\begin{flalign}
 C = \frac{2 \pi e^2}{A} \sum_{a} \frac{(\mathbf{k}/k) \cdot  Z_a^m \cdot (\mathbf{k}^a_{\text{LO}}/|\mathbf{k}^a_{\text{LO}}|) }{\sqrt{2M_a \omega_\text{LO} / \hbar}}
\end{flalign}
\end{subequations}
Here, $\varepsilon_{\text{i}}$ and $\varepsilon_{\text{o}}$ denotes the inorganic layer high-frequency dielectric constant and the organic layer high-frequency dielectric constant, respectively. We use high-frequency dielectric constants here since the exciton binding energies are much greater than the dominant LO phonon energies for the systems in this work\cite{blancon2018scaling}. Furthermore, the denominator of the full expression can be Taylor expanded to linear order in $k$ to yield
\begin{subequations} \label{s3}
\begin{flalign}
 |g(\mathbf{k})| \approx \frac{C}{\varepsilon_{\text{o}} + nr_{\text{eff}}k} 
\end{flalign}
where
\begin{flalign}
r_{\text{eff}} = \varepsilon_{\text{o}} d \bigg[ \frac{\varepsilon_{\text{o}}}{3 \varepsilon_{\text{i}}} + \frac{\varepsilon_{\text{i}}}{2 \varepsilon_{\text{o}}} \bigg( 1 - \bigg[ \frac{\varepsilon_{\text{o}}}{\varepsilon_{\text{i}}} \bigg]^2 \bigg) \bigg]  
\end{flalign}
\end{subequations}
For the majority of this work, since the dielectric constants of the layers differ considerably from each other, this approximate matrix element was found to be in good agreement with the full matrix element. Only in the case where the dielectric constants of layers become close to each other should it be necessary to use the full matrix element\cite{sohier2016two}.

\subsection{Getting rid of Born effective charges}
\noindent
A mass-weighted mode-effective Born effective charge tensor $Z^*_{v}$ was used in Ref.~\cite{sio2022unified} to replace $C$ in Eqs.\ref{s2} and \ref{s3} with 
\begin{equation} \label{s4}
C' = \frac{2\pi e^2}{A} Z^*_{v} \sqrt{\frac{\hbar}{2M_o \omega_{\text{LO}}}} 
\end{equation}
Here, $M_o$ is an arbitrary reference mass used to keep the mode-effective Born charge tensor unit-less. 
Then, we can deal with the Born effective charge tensor by utilizing the following expression (Eqs. 5.79 and 5.84 in \cite{dolgov1989dielectric})
\begin{equation} \label{s5}
\frac{(Z^*_v)^2}{\omega_{{LO},v}} = \frac{\Omega M_o \omega_{{LO},v}}{4 \pi e^2}(\varepsilon^\infty)^2 \bigg(\frac{1}{\varepsilon^\infty} - \frac{1}{\varepsilon^0} \bigg) 
\end{equation}
Here, $\Omega = A (nd + D)$ is the volume of a supercell consisting of one inorganic layer of length $nd$ and one organic layer of length $D$. We can decompose the overall dielectric constant $\varepsilon$ of the cell as
\begin{equation} \label{s6}
nd\varepsilon_{\text{i}}^{\infty,0} + D\varepsilon_{\text{o}}^{\infty,0} = (nd+D) \varepsilon^{\infty,0}  
\end{equation}
Then, we obtain
\begin{equation} \label{s8}
\frac{(Z^*_v)^2}{\omega_{{LO},v}} \approx \frac{ A  M_o \omega_{{LO},v}}{4 \pi e^2} \big[nd(\varepsilon_{\text{i}}^{0} -\varepsilon_{\text{i}}^{\infty}) \big] \bigg( \frac{nd\varepsilon_{\text{i}}^{\infty} + D \varepsilon_{\text{o}}^{\infty} }{nd\varepsilon_{\text{i}}^{0} + D \varepsilon_{\text{o}}^{0}}  \bigg) 
\end{equation}
Note that the organic layers in most  layered perovskites are fairly non-polarizable. In such a case, the static and high-frequency dielectric constants of those layers are approximately the same. Also, we revert back to the notation of the main text where $\varepsilon_{\text{i}} = \varepsilon_{\text{i}}^{\infty}$. Then, in the limit of a Rytova-Keldysh geometry ($D \gg nd$), we have that
\begin{equation} \label{s9}
\frac{(Z^*_v)^2}{\omega_{{LO},v}} \approx \frac{ A  M_o \omega_{{LO},v}}{4 \pi e^2} \big[nd(\varepsilon_{\text{i}}^{0} -\varepsilon_{\text{i}}) \big]  
\end{equation}
Plugging this back into the expression for $g(\mathbf{k})$ gives us
\begin{equation} \label{s10}
g(\mathbf{k}) = -i\frac{ \bigg[\pi e^2 nd \hbar \omega_{\text{LO}} (\varepsilon_{\text{i}}^0 - \varepsilon_{\text{i}}) /2A \bigg]^{1/2}} {\varepsilon_\text{o} +n  r_{\text{eff}}k}  
\end{equation}
Taking the magnitude of Eq.~\ref{s10} gives us Eq. 8a and 8b in the main text. 
\noindent
\subsection{Electron and Hole Path Integral Formalism}
\noindent
\textcolor{black}{With all of the terms of the model Hamiltonian being appropriately defined throughout the main text, the partition function, $\mathcal{Z}$, can be expressed as a path integral
\begin{equation} \label{part}
\mathcal{Z} = \int \mathcal{D}[ \boldsymbol{r}_e, \boldsymbol{r}_h, \boldsymbol{q}_{\boldsymbol{k}} ] \exp\bigg[-\frac{1}{\hbar}\mathcal{S}[\boldsymbol{r}_e, \boldsymbol{r}_h, \boldsymbol{q}_{\boldsymbol{k}}] \bigg]
\end{equation}
\textcolor{black}{where $\mathcal{D}[ \boldsymbol{r}_e, \boldsymbol{r}_h, \boldsymbol{q}_{\boldsymbol{k}} ]$ is the integration measure over all of the electron, hole, and phonon degrees of freedom.} The Euclidean action $\mathcal{S}$ is defined over an imaginary time $\tau$ as
\begin{equation} \label{totAct}
\mathcal{S} = \int_{0}^{\beta \hbar} d\tau \, \mathcal{H}_{\text{eh}, \tau} + \mathcal{H}_{\text{ph}, \tau} + \mathcal{H}_{\text{int}, \tau}  
\end{equation}
with $\beta = (k_\mathrm{B} T)^{-1}$ where $k_\mathrm{B}$ is Boltzmann's constant, and $T$ is the temperature of the system. Note that Eqs. \ref{part} and \ref{totAct} represent a classical isomorphism of the quantum Hamiltonian defined by Eq. 1 of the main text. Since the phonon degrees of freedom are coupled linearly to the electronic degrees of freedom, the phonon degrees of freedom can be integrated out to yield
\begin{equation} \label{phPartition}
\mathcal{Z} = \mathcal{Z}_{\text{ph}} \int \mathcal{D}[ \boldsymbol{r}_e, \boldsymbol{r}_h ] \exp \bigg[-\frac{1}{\hbar} \mathcal{S}_{\text{eff}} \bigg]
\end{equation}
where $\mathcal{Z}_{\text{ph}}$ is the phonon partition function. 
The action corresponding to the effective Hamiltonian can be written as an imaginary time influence functional\cite{park2022nonlocal,park2022renormalization}
\begin{eqnarray} \label{effAction}
\mathcal{S}_{\text{eff}} &=&\int_{\tau} \frac{m_e}{2}\dot{\mathbf{r}}_e^2 +\frac{m_h}{2}\dot{\mathbf{r}}_h^2 + V(\mathbf{r}_e,\mathbf{r}_h) \\
&-& \alpha \hbar \omega_{\text{LO}}^2 \sum_{i,j\in e,h}   \Gamma_{i,j}  \:  \int_{\tau} \int_{\tau '} e^{-\omega_{\text{LO}}|\tau - \tau '|} f(|\textbf{r}_{i,\tau}-\textbf{r}_{j,\tau '}|)\nonumber
\end{eqnarray}
where $\Gamma_{i,j}=1$ for $i=j$ and $-1$ otherwise, while
\begin{equation} \label{alpha}
\alpha = \frac{ e^2 nd}{16 n^2 r_{\text{eff}}^2 \hbar \omega_{\text{LO}}  }   (\varepsilon_\mathrm{i}^0 - \varepsilon_\mathrm{i} )
\end{equation}
is a dimensionless coupling parameter and $f(|\mathbf{r}_{i,\tau}-\mathbf{r}_{j,\tau '}|)$ is defined by
\begin{equation} \label{kernel}
f(r)=    \int_{\mathbf{k}} \frac{e^{i\mathbf{k} \cdot \mathbf{r} }} {(\varepsilon_{\text{o}} + n r_{\text{eff}} k )^2} 
\end{equation}
We can further deal with the $k$ space integral in Eq.~\ref{kernel}
\begin{equation} \label{s14}
\int_{\mathbf{k}} \frac{e^{i\mathbf{k} \cdot \mathbf{r}}} {(\varepsilon_{\text{o}} + nr_{\text{eff}} k )^2} = \bigg(\frac{A}{2 \pi} \bigg) \bigg(\frac{1}{\varepsilon_{\text{o}} ^2} \bigg) \int_{0}^{\infty} dk \frac{k\ J_o\big(k r\big)}{(1 + \frac{nr_{\text{eff}}}{\varepsilon_{\text{o}}}k)^2}
\end{equation}
Then, we obtain
\begin{equation}
\label{s16}
f(r) =   \bigg(\frac{A}{4 \pi n^2 r_{\text{eff}}^2} \bigg)  \bigg[ \frac{2 \varepsilon_{\text{o}}}{nr_{\text{eff}}}r + \pi \bigg(  H_0 ( \frac{ \varepsilon_{\text{o}}}{nr_{\text{eff}}} r ) -  Y_0 ( \frac{ \varepsilon_{\text{o}}}{nr_{\text{eff}}} r ) \bigg) 
- \frac{\pi \varepsilon_{\text{o}}}{nr_{\text{eff}}} r  \bigg( H_1 ( \frac{ \varepsilon_{\text{o}}}{nr_{\text{eff}}} r ) - Y_1 ( \frac{ \varepsilon_{\text{o}}}{nr_{\text{eff}}} r ) \bigg)    \bigg]  
\end{equation}
\textcolor{black}{ Since we are concerned with zero temperature exciton binding properties, we additionally note  Eq. \ref{effAction} employs the zero temperature phonon correlation function. One could use the finite temperature phonon correlation function to capture the temperature dependence of exciton binding.}}

\newpage
\section{Additional Simulation Details}
\noindent 
In a path integral framework, our overall action after integrating out the phonon degrees of freedom is $\mathcal{S} = \mathcal{S}_{\text{eh}} + \mathcal{S}_{\text{int}}$. We first deal with $\mathcal{S}_{\text{eh}}$. After discretizing this action into $N$ imaginary time slices, we have
\begin{equation} \label{s18}
\mathcal{S}_{\text{eh}} = \sum_{i = 1}^{N} \frac{m_{\text{e}}N}{2 \beta^2 \hbar^2} ( \mathbf{r}_{{\text{e}},i+1  } - \mathbf{r}_{{\text{e}},i  } )^2 + \sum_{i = 1}^{N} \frac{m_{\text{h}}N}{2 \beta^2 \hbar^2} ( \mathbf{r}_{{\text{h}},i+1  } - \mathbf{r}_{{\text{h}},i  } )^2 + \frac{1}{N}\sum_{i = 1}^{N} V(r_i)
\end{equation}
where $\beta = (k_\mathrm{B} T)^{-1}$ is the inverse temperature, $m_{\text{e}}$ and $m_{\text{h}}$ are the masses of the electron and hole, and $\mathbf{r}_{{\text{e}},i  }$ and $\mathbf{r}_{{\text{h}},i  }$ are the positions of the electron and hole at imaginary time $i$. Furthermore, the electrostatic potential $V$ is defined by Eq.7 in the main text, and $r_i = |\mathbf{r}_{{\text{e}},i  } - \mathbf{r}_{{\text{h}},i  }|$ . To deal with the divergence at $r_{i} = 0$, we make the electrostatic potential a function of $(r_c^2 + r^2)^{1/2}$ rather than sole functions of $r$. The parameter $r_c$ is chosen such that the bandgap of the BTA and PEA perovskites are reproduced at $r = 0$ \cite{schnitker1987electron,park2022nonlocal,park2022renormalization}. For simplicity, we use $r_c$ values corresponding to the $n = 1$ bandgaps of the BTA\cite{cho2019optical} and PEA\cite{straus2022photophysics} perovskites for larger values of $n$. To validate the accuracy of this procedure, we find that numerical solutions for the Schrödinger equation for the Hamiltonian corresponding to $\mathcal{S}_{\text{eh}}$ agree well with the corresponding path integral Monte Carlo results. The values of $r_c$ used in this work are summarized in Table \ref{tbtestTable}.
\newline
\noindent
Lastly, in imaginary-time discretized form, the exciton-phonon action becomes
\begin{equation} \label{s19}
 \mathcal{S}_{\text{int}} = -  \frac{\alpha \beta \hbar^2 \omega_{\text{LO}}^2}{N^2} \sum_{i,j\in e,h}   \Gamma_{ij}  \  \sum_{s \neq t} e^{-\frac{\beta \hbar \omega_{\text{LO}}}{N}|s - t|} f(|\mathbf{r}_{i,s}-\mathbf{r}_{j,t}|)
\end{equation}
where $s,t \in [1,N]$.
\newline
\noindent
For the simulations, we find that $N = 200$ time slices provides sufficient convergence for the results obtained in this work. We run the simulations for between 2 and 5 million Monte Carlo sweeps and keep the acceptance rate to around 40$\%$ to ensure equilibration and sufficient statistics. For an efficient implementation of the kernel $f$ defined by Eq.~\ref{s16} and \ref{s19}, we use highly accurate polynomial approximations of Struve and Bessel functions.\cite{newman1984approximations} It should also be noted that the solutions to Poisson's equation for both geometries in Eq. 5 of the main text can only be expressed analytically in momentum space \cite{muljarov1995excitons,rytova2020screenedpotentialpointcharge,keldysh2024coulomb,barrera101978point}. Then, we must numerically Fourier transform these solutions into position space in order to use them for simulation. 
\newline
\noindent
We first describe the Fourier transform for the thin film geometry. For this case, the solution to Poisson's equation (Eq.3-5 of the main text) can be written as\cite{rytova2020screenedpotentialpointcharge}
\begin{equation} \label{keld}
 -eV(k,z_{\text{e}},z_{\text{h}}) = -\frac{2 \pi e^2}{\varepsilon_{\text{i}}}\bigg(e^{-k|z_{\text{e}}-z_{\text{h}}|} + \frac{2\delta}{e^{2knd}-\delta^2} \big[\delta \cosh \big(k(z_{\text{e}}-z_{\text{h}})\big) + e^{knd} \cosh \big(k(z_{\text{e}}+z_{\text{h}} - nd)\big) \big]   \bigg)   
\end{equation}
where $\delta$ is defined in Eq.~\ref{s2} and $z_{\text{e}},z_{\text{h}} \in [0,nd]$. We can now write a momentum space analogue to Eq.5 of the main text
\begin{equation}\label{kPot}
V(k) = -e\int_{z_{\text{e}}} \int_{z_{\text{h}}} P(z_{\text{e}}) P(z_{\text{h}}) V(k,z_{\text{e}},z_{\text{h}}) dz_{\text{h}} dz_{\text{e}} 
\end{equation}
Then, we can use particle-in-a-box densities 
\begin{equation} \label{Pz}
P(z_i) = \frac{2}{nd} \sin^2 \big(\frac{\pi z_i}{nd} \big)
\end{equation}
where $i \in \{e, h \}$, to write a quantum-confined thin film potential
\begin{equation} \label{keldAveraged}
V(k) = -\frac{2\pi e^2 } {k \varepsilon_{\text{i}}[(knd)^2 + 4\pi^2]^2} \bigg[3(knd)^3 + 20 \pi^2 (knd) +  \frac{32\pi^4}{(knd)^2 }\bigg( \frac{1-\delta(1+knd) + e^{knd}(\delta + knd - 1)}{(e^{knd} - \delta)} \bigg) \bigg]
\end{equation}
Finally, we can Fourier transform $V(k)$ into $r$-space through the following integral
\begin{equation} \label{fourierTrans}
V(r) = \frac{1}{2\pi} \int_{0}^{\infty}k V(k) J_{0}(kr) dk
\end{equation}
The provided source code evaluates this integral for the thin film geometry through the Hankel transform library\cite{murray2019hankel,rohit_rana_2024_12683301}.
\newline
\noindent
We use th same Fourier transform procedure for the alternating slab, crystal geometry. In this case, the solution to Poisson's equation in momentum space is written as\cite{muljarov1995excitons}
\begin{subequations}\label{muljarov}
\begin{multline}
-eV(k,z_{\text{h}},z_{\text{h}}) = \frac{2\pi e^2}{\varepsilon_{\text{i}}k} \sinh \big(k|z_{\text{e}} - z_{\text{h}}|\big) - \frac{2\pi e^2}{\varepsilon_{\text{i}}k \sinh \big(k_o\big)} \bigg[ 2st \sinh \big(kD\big) \cosh\big(k(z_{\text{e}} + z_{\text{h}} + nd) \big) \bigg] \\
-\frac{2\pi e^2}{\varepsilon_{\text{i}}k \sinh \big(k_o\big)} \cosh\big(k(z_{\text{e}} - z_{\text{h}}) \big) \bigg[s^2 \sinh\big(k(nd+D) \big) + t^2 \sinh\big(k(D-nd) \big)  \bigg]  
\end{multline}
where
\begin{flalign}
s = \frac{1}{2}(1+\varepsilon_{\text{o}}/\varepsilon_{\text{i}}), \: t =  \frac{1}{2}(1-\varepsilon_{\text{o}}/\varepsilon_{\text{i}})
\end{flalign}
and
\begin{flalign}
\frac{\varepsilon_{\text{o}}}{\varepsilon_{\text{i}}}\cosh \big( k_o \big) = s^2 \cosh \big( k(nd+D) \big) - t^2 \cosh \big( k(D-nd) \big)
\end{flalign}
\end{subequations}
When averaging out the $z$-dependencies from Eq.~\ref{muljarov}, we use $P(z_i - nd)$ in Eq.~\ref{Pz} to get a quantum-confined $V(k)$ similar to Eq.~\ref{keldAveraged}. However, we highlight that trying to use this resulting $V(k)$ in Eq.~\ref{fourierTrans} is problematic due to the highly oscillatory nature of the integrand. This issue was resolved by Hansen et al.\cite{hansen2023mechanistic} by setting the upper bound of the integral in Eq.~\ref{fourierTrans} as the value of $k$ that maximizes the amplitude of the integrand for a given value of $r$. We then used the openly available code provided by these authors to conduct this procedure along with particle-in-a-box densities and the parameters listed in this work to get $V(r)$ for the crystal geometry.
\newpage
\section{Sensitivity to Static Dielectric Constant}
\noindent
To probe the sensitivity of the exciton binding energy of our model as a function of static dielectric constant, we studied the $n=1$ case of the thin-film geometry for the BTA and PEA ligands. We computed exciton binding energies using values of $\varepsilon_{i}^{0} =$ 13.0 (2D/Main Text), 18.6, and 24.1 (Bulk)\cite{park2022renormalization} for $n=1$. As seen from Fig.\ref{static}, the sensitivity of the exciton binding energies with respect to the static dielectric constant becomes more modest if the dielectric mismatch between inorganic and organic layers is greater. 
\section{Derivation of Variational Bound}
\noindent
We start by considering a free electron trial action
\begin{equation} \label{s20}
\mathcal{S}_o = \int_{0}^{\beta \hbar} d\tau \frac{m_{\text{e}}}{2} \dot{\mathbf{r}}^2_{\text{e}} 
\end{equation}
where $\dot{\mathbf{r}}_{\text{e}} = \dot{\mathbf{r}}_{\text{e}}(\tau)$ represents the electron degrees of freedom at imaginary time $\tau$. Then, the Feynmann-Bogolyubov inequality\cite{feynman1962mobility} yields
\begin{equation} \label{s21}
\Delta E \geq -E_{o} - \frac{1}{\beta} \langle \mathcal{S}_{\text{e}} - \mathcal{S}_o  \rangle_o
\end{equation}
where $E_o$ is the energy associated with the trial action, $\mathcal{S}_{\text{e}}$ is the action of a single electron interacting with the lattice through a thin-film Fröhlich interaction, and the expectation value is evaluated with respect to the distribution associated with the trial action. Then, the polaron binding energy is given in the low-temperature limit as
\begin{equation} \label{s22}
\Delta E \geq \lim_{k_\mathrm{B} T \to 0} k_\mathrm{B} T \bigg<  \alpha \hbar \omega_{\text{LO}}^2 \int_{0}^{\beta \hbar} d \tau \int_{0}^{\beta \hbar} d \tau' e^{-\omega_{\text{LO}} |\tau - \tau'|} f(|\mathbf{r_\tau} - \mathbf{r_{\tau'}}|) \bigg>_o  
\end{equation}
We will first deal with the expectation value of $f$
\begin{equation}
 \big < f(|\mathbf{r_\tau} - \mathbf{r_{\tau'}}|) \big>_o = \frac{4\pi n^2 r_{\text{eff}}^2 }{A}\sum_{k} \frac{1}{(\varepsilon_{\text{o}} + nr_{\text{eff}}k)^2} \big< e^{ik \cdot |r_\tau - r_{\tau'} |} \big>_o = \frac{4\pi n^2 r_{\text{eff}}^2 }{A \hbar} \sum_{k} \frac{1}{(\varepsilon_{\text{o}} + nr_{\text{eff}}k)^2}  e^{-\hbar k^2 |\tau - \tau'|/(2m_{\text{e}})}
 \end{equation}
Our inequality now becomes
\begin{equation} \label{s24}
\begin{split}
\Delta E \geq \lim_{k_\mathrm{B} T \to 0} k_\mathrm{B} T \bigg[  \frac{4\pi n^2 r_{\text{eff}}^2 \alpha  \omega_{\text{LO}}^2 }{A } \sum_{k} \frac{1}{(\varepsilon_{\text{o}} + nr_{\text{eff}}k)^2}  \int_{0}^{\beta \hbar} d \tau \int_{0}^{\beta \hbar} d \tau' e^{-[\omega_{\text{LO}} + \hbar k^2 /(2m_{\text{e}}) ]|\tau - \tau'|} \bigg]\\   
= \frac{4\pi n^2 r_{\text{eff}}^2 \alpha  \omega_{\text{LO}}^2 m_{\text{h}} }{A } \sum_{k} \frac{1}{(\varepsilon_{\text{o}} + nr_{\text{eff}}k)^2} \cdot \frac{4\hbar}{\hbar k^2 + 2m_{\text{e}} \omega_{\text{LO}}}\\
\end{split}
\end{equation}
Due to the spacing between successive $\mathbf{k}$, the sum can then be replaced by an integral
\begin{equation} \label{s25}
 \frac{4\pi n^2 r_{\text{eff}}^2 \alpha  \omega_{\text{LO}}^2 m_{\text{h}} }{A } \cdot \frac{A}{(2\pi)^2} \int_{0}^{2\pi} \int_{0} dk \ d\phi \frac{4\hbar k}{(\hbar k^2 + 2m_{\text{e}} \omega_{\text{LO}}) (\varepsilon_{\text{o}} + nr_{\text{eff}}k)^2 } \\
\end{equation}
Evaluation of the integrand ultimately yields
\begin{equation} \label{s26}
\Delta E \geq \alpha  \hbar \omega_{\text{LO}} \xi \left( \frac{n r_{\text{eff}}}{\varepsilon_\text{o}} \Bigg/\sqrt{\frac{\hbar}{2 m_{\text{e}} \omega_{\text{LO} }}} \right )
\end{equation}
where
\begin{equation} \label{s28}
\xi (a) = 4\frac{ (a^4-a^2)       \ln \big(a\big) + \pi a^3  - a^2 - a^4          }{ (1 + a^2)^2}
\end{equation}
depends on the ratio of the dielectric confinement length $n r_{\text{eff}}/\varepsilon_\text{o}$ to the polaron radius $\sqrt{\hbar/2 m_{\text{e}} \omega_{\text{LO} }}$.
$\Delta E$ is plotted as a function of $n$ for the BTA and PEA perovskites in Fig.\ref{figvar}. Even though the quantitative agreement with Fig. 2a of the main text can be improved upon with a better trial action, the bound derived in this work captures a non-monotonic behavior that qualitatively resembles the simulation results.

\section{Repository}
\noindent
The source code for the Monte Carlo calculations, and the
code used to generate the figures are openly available online.\cite{rohit_rana_2024_12683301}
\newpage

\begin{center}
\begin{tabular}{lrrrrr} 
 \hline
  & BTA/Film &  \ BTA/Crystal &\ PEA/Film &\ PEA/Crystal \\  
 \hline\hline
 $r_c$ (au) &  0.918 & 1.357 & 0.898  & 1.401 \\
 Band Gap (meV) &  2.80 & 2.80 & 2.58  & 2.58 \\
 \hline
\end{tabular}
\captionof{table}{Pseudopotential parameters $r_c$ and bandgaps (Eq.~\ref{s18}).}
\label{tbtestTable}
\end{center}
\newpage
\begin{center} \label{S1}
  \includegraphics[scale=0.45]{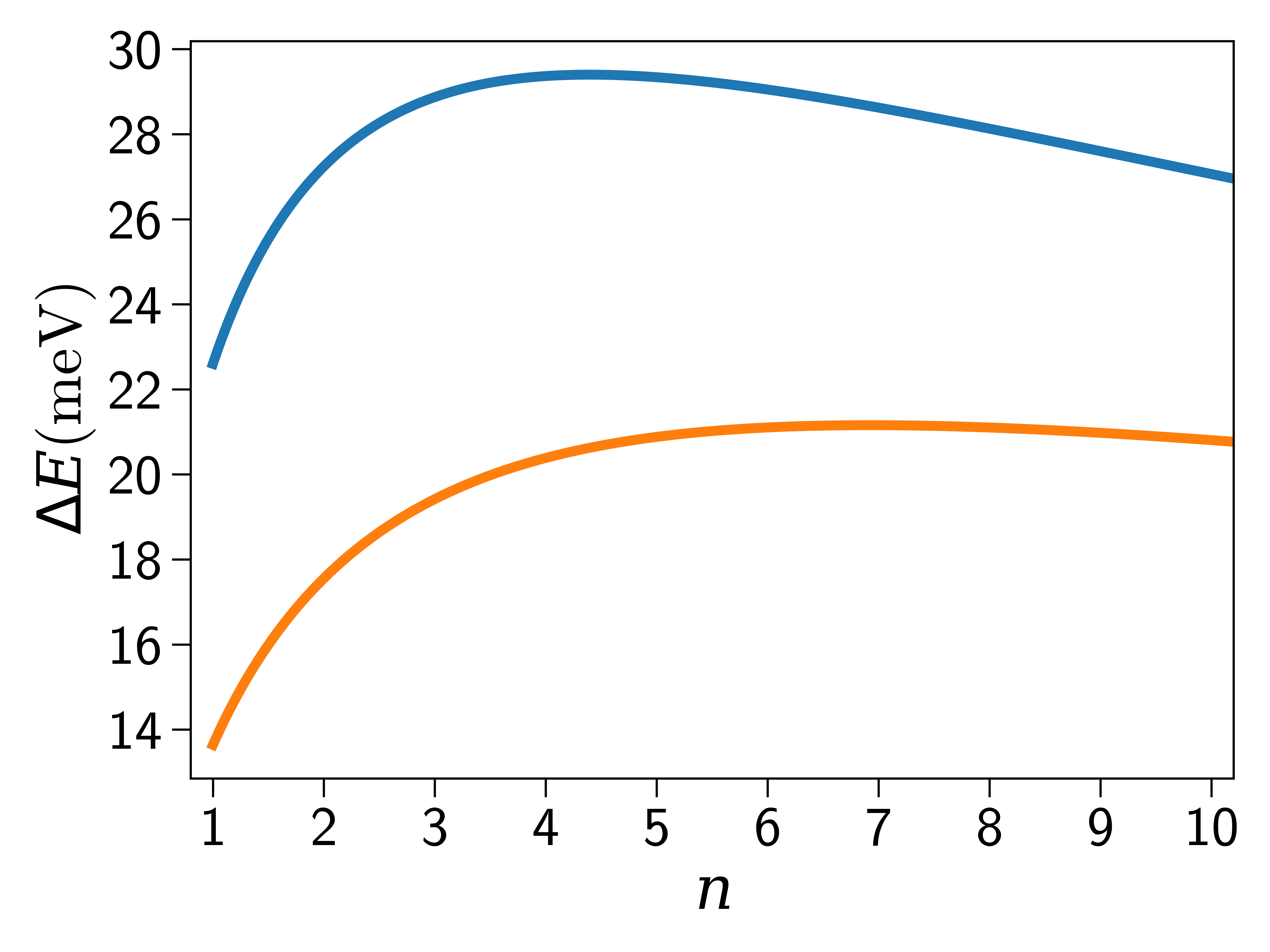}
  \captionof{figure}{Variational Bound (Eq.~\ref{s26} and \ref{s28}) as a function of inorganic well thickness. Top line and bottom line correspond to BTA and PEA perovskites, respectively.}
  \label{figvar}
\end{center}
\newpage
\newpage
\begin{center} \label{S2}
  \includegraphics[scale=0.45]{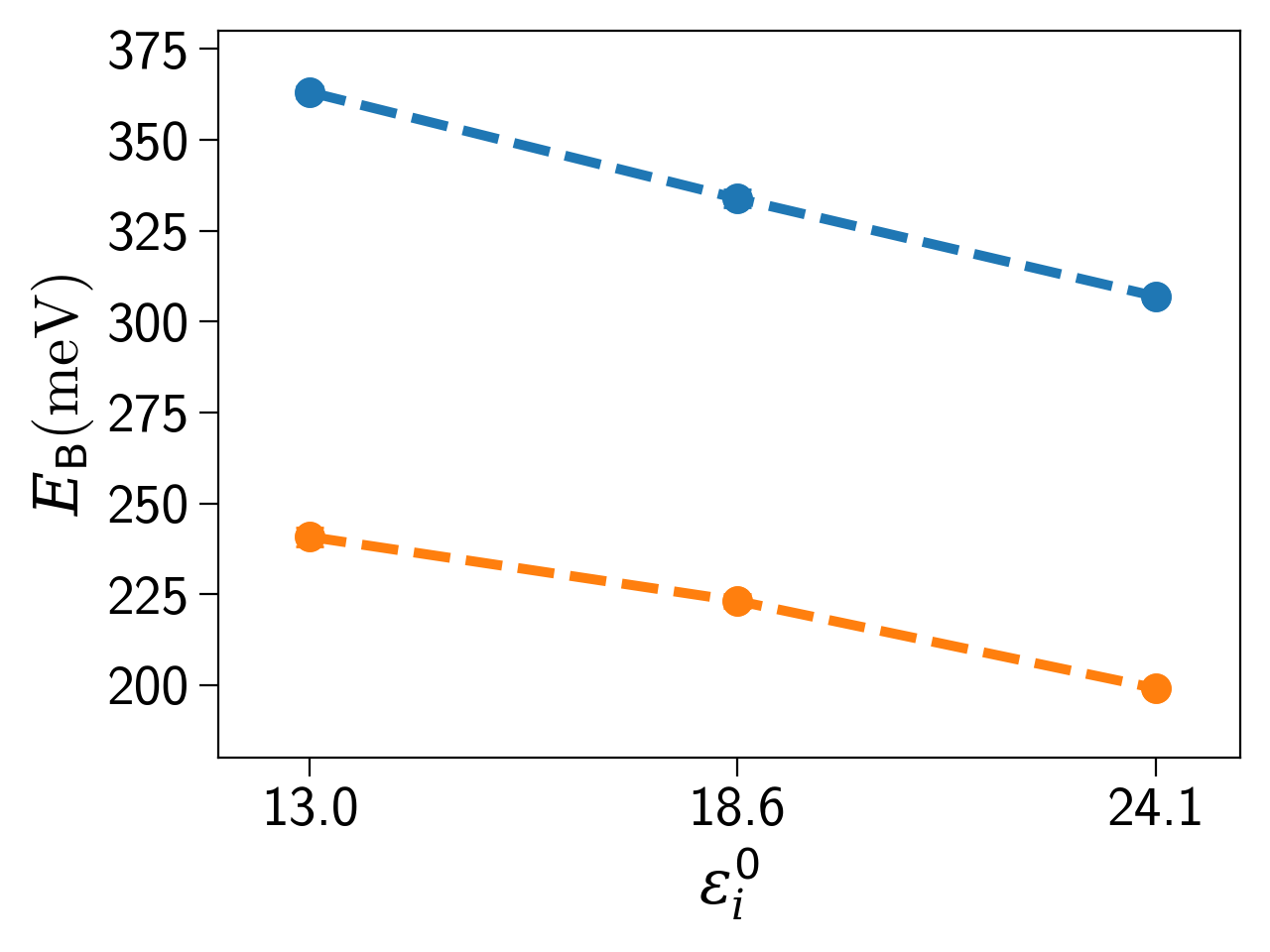}
  \captionof{figure}{Exciton binding energy as a function of inorganic well static dielectric constant $\varepsilon_{i}^{0}$ for $n=1$. Top line and bottom line correspond to BTA and PEA perovskites, respectively.}
  \label{static}
\end{center}
\newpage
\bibliography{CITE_SI}